\documentclass[onecolumn,preprintnumbers,amsmath,amssymb]{revtex4}
\usepackage{graphicx}% Include figure files
\usepackage{dcolumn}% Align table columns on decimal point
\usepackage{bm}% bold math
\usepackage{times}

\begin{document}

\title{Near-edge x-ray absorption fine structure investigation of graphene}
% Force line breaks with \\

\author{D. Pacil\'e$^{1, 4, 5}$}
\author{M. Papagno$^{1}$}
\author{A. Fraile Rodr\'{\i}guez$^{2}$}
\author{M. Grioni$^{3}$}
\author{L. Papagno $^{1}$}%
\affiliation{%
$^{1}$~\mbox{Istitut. Naz. di Fisica Nucleare (INFN) and Dip. di Fisica, Universit$\grave{a}$ della Calabria, 87036 
\newline
Arcavacata di Rende, Cosenza, Italy} \\
$^{2}$~\mbox{Swiss Light Source, Paul Sherrer Institut, 5232 Villigen PSI, Switzerland}\\
$^{3}$~\mbox{Ecole Polytechnique F\'ed\'ederale de Lausanne, Institut de Physique des Nanostructures, CH-1015 Lausanne, Switzerland}\\
}%

\author{\c{C}. \"{O}. Girit$^{4, 5}$}
\author{J. C. Meyer $^{4, 5}$}
\author{G. E. Begtrup $^{4, 5}$}
\author{ A. Zettl $^{4, 5}$}%

\affiliation{%
$^{4}$~\mbox{Department of Physics, University  of California, Berkeley, CA 94720, USA }\\
$^{5}$~\mbox{Materials Sciences Division, Lawrence Berkeley National Laboratory, Berkeley, CA 94720, USA }\\
}%

\date{\today}
\begin{abstract}
\noindent
We report the near-edge x-ray absorption fine structure (NEXAFS) spectrum of a single layer of graphite (graphene) obtained by micromechanical cleavage of Highly Ordered Pyrolytic Graphite (HOPG) on a SiO$_{2}$ substrate. We utilized a PhotoEmission Electron Microscope (PEEM) to separately study single- double- and
few-layers graphene (FLG) samples. In single-layer graphene we observe a splitting of the  $\pi*$ resonance and a clear signature of the predicted  $interlayer$ state. The NEXAFS data illustrate the rapid evolution of the 
electronic structure with the increased number of layers.

\end{abstract}

\maketitle

The recent discovery of a single sheet of graphite  \cite{Novoselov}, called graphene, has opened up a new area of condensed matter physics. Graphene proves that materials just one atom thick may exist, with exciting prospects for applications.  Its unusual electronic spectrum, where charge carriers mimic massless relativistic particles  \cite{Novoselov2, Zhang}, also provides an unexpected bridge between condensed matter physics and quantum electrodynamics.

The method to obtain single sheets of graphite  \cite{Novoselov}, called micro-mechanical cleavage,  allows easy production of sample with a typical size of few tens of microns, ideal for ballistic transport and Quantum Hall effect measurements, but inappropriate for many conventional spectroscopy investigations in Ultra High Vacuum (UHV) conditions. 
In the absence of new and more efficient ways to make graphene, samples obtained by micro-mechanical cleavage of bulk graphite are used in a limited class of experiments, where the size and the identification of thin flakes is possible. Indeed, after the cleavage with simple adhesive tape, graphene crystallites left on the SiO$_{2}$ substrate are extremely rare and hidden amongst hundreds of thicker flakes. Conventional surface science probes of the electronic and structural properties of materials, are then ruled out, unless they are coupled to a microscope. On the other hand, single- and few-layers graphene (FLG) samples have been grown epitaxially by chemical vapour deposition of hydrocarbons on metal substrates \cite{Oshima} and by thermal decomposition of SiC  \cite{ Forbeaux}. In both cases, the hybridization of graphene with the substrate is an unavoidable complication, although graphene on SiC preserves most of the electronic properties expected for a free layer \cite{ Zhou, Ohta, Bostwick}.

In this Letter we report the near-edge x-ray absorption fine structure (NEXAFS) spectra of a free layer of graphene,
and of few-layers graphene (FLG) samples, obtained by a PhotoEmission Electron Microscope (PEEM) in UHV conditions.
The spectrum of graphene exhibits  a new structure  below the $\pi*$ resonance, reflecting its peculiar density of states (DOS) above the Fermi level \cite{Trickey}, and a peak between the  $\pi*$  and $\sigma*$ resonances. We attribute the latter to the analog of the $interlayer$ state of graphite, which was predicted to exist even in single-layer graphene \cite{ Fischer, Posternak}. By increasing the number of layers, the electronic structure rapidly evolves towards that of bulk graphite, under the influence of the weak interlayer interaction \cite{Latil, Partoens}.

Our samples were prepared by micro-mechanical cleavage of HOPG on SiO$_{2}$ substrates and characterized by  optical microscopy (OM) and Raman spectroscopy, to identify single layers of graphite and thicker flakes \cite{Ni, Ferrari}.
The laterally resolved NEXAFS experiments were carried out at the SIM beamline \cite{ Quitmann} of the Swiss Light Source, using an Elmitec PEEM equipped with an energy analyzer. The image contrast in PEEM can arise from several sources including topography, element specificity, and chemical bonding in the sample. In order to obtain element-specific PEEM contrast, images collected at the peak of the C 1s-absorption absorption edge (285.5 eV) were normalized by corresponding images measured below the absorption edge (282.5 eV). This procedure enhances the elemental contrast and reduces topographic contrast and illumination inhomogeneities. 

%\begin{figure}
%\includegraphics[width=3.2in]{Fig1}% Here is how to import EPS art
%\caption{\label{Fig1:epsart} 
%OM images [(a), (c) and (e)] and PEEM ones [(b), (d) and (f)] of some selected samples (S1, S2 and S3). Scale bars are: 5 $\mu$m in (a) and (e); 10 $\mu$m in (c). }
%\end{figure}

%\begin{figure}
%\includegraphics[width=3.4in]{Fig2}% Here is how to import EPS art
%\caption{\label{Fig2:epsart} 
%C(K)-edge photoabsorption spectra of (from the bottom): graphene, bilayer graphene and FLG samples. Dashed lines show the C1s -$\pi*$ and C1s-$\sigma*$ transitions.}
%\end{figure}

Fig. 1 shows a direct comparison between OM images [(a), (c) and (e)] and PEEM ones [(b), (d) and (f)] of some selected samples (S1, S2 and S3). Sample S3 was prepared using a different technique: trenches 2$\mu$m wide were patterned on the SiO$_{2}$/Si substrate by photolithography and then etched to a depth of 150 nm. After the cleavage procedure, single sheets suspended over trenches provided free-standing graphene membranes. PEEM images were collected by setting the photon energy just above the C 1s edge, for a total exposure time of about 30 s. The contrast provided by the PEEM is surprisingly high, even for a single layer, demonstrating the possibility to identify and analyze the electronic properties of graphene in a UHV experiment. Moreover, different thickness of the flakes, quantified by means of the OM contrast method \cite{Ni}, can be easily recognized in the PEEM images. 

%\begin{figure}
%\includegraphics[width=2.4in]{Fig3}% Here is how to import EPS art
%\caption{\label{fig3:epsart} 
%Expanded C(K)-edge photoabsorption spectra of graphene and bilayer graphene extracted from Fig. 2. The inset shows corresponding DOS calculations, extracted from Ref. \cite{Trickey}, in the energy region (0.0$\div$5.0) eV above $E_{F}$. Arrows show the double structure of the $\pi*$ resonance in graphene. The separation between the two features in the theoretical DOS of 1L
%is about 0.8 eV, while the experimental separation is between 1.2 and
%1.8 eV for our samples.}
%\end{figure}

%\begin{figure}
%\includegraphics[width=3.2in]{Fig4}% Here is how to import EPS art
%\caption{\label{fig4:epsart} 
%C(K)-edge photoabsorption spectra of graphene taken from sample S2. In the inset, the Raman spectrum of graphene at 514 nm  showing the single component of the 2D peak. }
%\end{figure}

Photoabsorption spectra were extracted by processing a full package of PEEM images obtained by tuning the photon energy across the C 1s edge, and recording the total electron yield (TEY) from the region of interest. In order to normalize the C 1s absorption spectra, the images were divided by the x-ray absorption from an area outside the region of interest.  All measurements were performed at grazing incidence (16$^{\circ}$) with a linear polarization almost perpendicular to the basal plane of graphite, a configuration ($\textbf{E}$$\|$$\textbf{c}$) which enhances transitions into final states of $\pi$ symmetry \cite{Rosenberg}. Fig.2 shows C 1s absorption spectra obtained for graphene, bilayer graphene and FLG samples. The main peak at 285.5 eV ($\pi*$ resonance), is associated with the conduction $\pi$ states around the M and L points of the Brillouin zone (BZ) \cite{ Fischer}. The structure at 291.5 eV is due to the dispersionless $\sigma$ states at the $\Gamma$ point of the BZ. Higher energy features are due to transitions towards higher-lying states of  $\pi$ or $\sigma$ symmetry  \cite{Rosenberg}. They are better defined in thicker samples (above 5 layers), where the electronic structure is mainly reminiscent of the graphite band structure.  The electronic structure of graphite exhibits strong modifications for a finite number of layers, although the interlayer coupling between its planes is supposed to be very weak. These changes produce corresponding drastic changes of the transport properties   \cite{Kopelevich}.  Graphene is a zero-gap semiconductor with a linear Dirac-like spectrum around the Fermi energy ($E_{F}$), while graphite is a semimetal with a band overlap of 41 meV  \cite{Partoens}. In the intermediate region, bilayer graphene double $\pi$ states show parabolic dispersion at $E_{F}$, with an overlap of about 1.6 meV due to the interaction between B and B$^{\prime}$ carbon atoms of different planes. FLG show different parabolic-like $\pi$ states, with an increasing  band overlap as a function of the number of layers, up to that of graphite above ten layers. 

In a band-theory approach \cite{Muller}, the near-edge structure is the sum of the transition rates to all possible unoccupied one-electron states. Each transition rate is further separated into two components: the matrix element between the initial and the final states, and the DOS projected along the k$_{z}$ direction. Both terms vary with the number of graphene layers. Moreover, matrix elements strongly depend on polarization. In graphite, the states above the  $\pi*$ resonance have mainly $\sigma$ symmetry \cite{Rosenberg}. Their contribution to the NEXAFS spectrum is strongly reduced in our experimental set-up ($\textbf{E}$$\|$$\textbf{c}$) and this explains why the spectra of Fig. 2 are dominated by the first resonance. The absence of those high energy features is particularly evident in the photoabsorption spectrum of graphene (Fig.2, Fig. 4), where, in addition, the number of absorbing atoms per unit area is reduced to its minimum value.
Near-edge structures can also be interpreted by multiple-scattering (MS) theory \cite{Weng}, which considers the scattering of the excited electron wave function by the neighboring atoms. The presence of a $cage$ around the absorbing atom is reflected into the features of the near-edge structure, thus explaining the absence of signal for a single layer (Fig. 2, Fig. 4) when the polarization of the light would select neighboring atoms above and below the plane.

Fig. 3 shows a close up around the $\pi*$ resonance of the graphene and bilayer graphene spectra extracted from Fig.2. Here, a clear peak located below the $\pi*$ resonance, at about 283.7 eV, can be seen for graphene, while a broader shoulder is present for bilayer graphene. A similar structure was  observed in NEXAFS measurements performed on nanographite grains growth on Pt(111), and explained in term of an edge-derived electronic state  \cite{Entani}. Specifically, it was assigned to grains with zigzag edges, for which edges states are expected just above the Fermi level, in contrast with grains with armchair edges \cite{Nakada}. Our measurements show a similar structure, but the edge contribution is negligible considering the size ($>$ 10 $\mu$m) of our samples. On the other hand,  the DOS (inset of Fig.3) of 1 and 2 layers of graphene, calculated in the Local-Density Approximation (LDA) by S. B. Trickey $et al.$  \cite{Trickey}, is in good agreement with the experimental data. Considering the energy resolution of our spectra ($\sim$ 0.2 eV), the four peaks (starting 1 eV above the Fermi level) in the DOS of bilayer graphene will yield a shoulder below the $\pi*$ resonance. On the other hand, two pronounced peaks are present in the DOS of graphene, which are clearly resolved in our spectra. This double structure of the $\pi*$ resonance is associated with two zero-slope points  along the  $\overline{\mbox{M}}\overline{\mbox{K}}$ high-symmetry direction in the band structure of graphene \cite{Trickey}. Therefore, the peak at 283.7 eV cannot be related to a zigzag edge effect of graphene, but must be attributed to the peculiar unoccupied DOS of a single layer of carbon atoms.

Another interesting aspect in the NEXAFS spectrum of graphene, as well as of FLG, is a feature located at about 288 eV (Fig.2), between the $\pi*$ and $\sigma*$ resonances. This feature can be clearly seen in the spectrum of Fig. 4, obtained from the graphene flake of sample S2, where a higher spatial integration was performed. Considering the high inertness of graphene flakes, possible contamination of C-H species at surface, which would give $\sigma*$(C-H) transitions located between the $\pi*$ and $\sigma*$ resonances \cite{Kikuma}, are not expected to make a significant contribution. This was also confirmed by Raman spectra taken on the same samples. The inset of Fig. 4, indeed, shows the Raman spectrum of sample S2 taken at 514nm and centered on the sharp 2D band, confirming for this region the thickness of one layer \cite{ Ferrari} and showing no structure related to $ \nu$(C-H) vibrations at about 2900 cm$^{-1}$ \cite{B. Zhang}. In addition, extreme care was taken to avoid sample damage or beam induced heating during the NEXAFS experiment. 

The peak located at about 288 eV, clearly defined in the photoabsorption spectrum of Fig.4, can be ascribed to a graphene analog of the $interlayer$ state of graphite. This state has been the subject of theoretical and experimental investigations in the past, as it changed the commonly accepted description of the graphite band structure. Indeed, beside $\sigma$ and $\pi$ bonding states, followed by $\pi*$ and $\sigma*$ antibonding states  above the Fermi level, theory predicted an additional conduction band of three dimensional character, namely showing strong dispersion along the perpendicular direction \cite{ Posternak}.  This state was called $interlayer$ state because its charge density is mainly confined between the basal planes, although it was believed to exist for a single layer of atoms as well. It helped to understand the electronic behavior of alkali-metal Graphite Intercalation Compounds (AGICs) \cite{Posternak, Csanyi}, demonstrating that no additional state arises from the alkali s electrons, and that the free-electron band of AGICs preexists above the Fermi level in pure graphite. The  $interlayer$ state was first observed in HOPG by angle-resolved inverse photoemission \cite{ Fauster} and  NEXAFS  \cite{ Fischer}.  Nevertheless, the possibility to detect this state by C 1s TEY spectra of pure HOPG was questioned  \cite{Kurmaev}, and its location with respect to the Fermi level was also a source of controversy \cite{Batson}. The present results clearly show  a structure between the $\pi*$ and $\sigma*$ resonances, similar to NEXAFS data on HOPG \cite{ Fischer}, and establish the existence of the $interlayer$ state in graphene. 
These observations were confirmed also for sample S3, showing that graphene flakes deposited on SiO$_{2}$ do behave, with respect to the electronic properties probed by a NEXAFS investigation, like suspended membranes.

In conclusion, laterally resolved absorption spectroscopy performed in UHV conditions by a PEEM microscope has allowed us to establish the C 1s absorption spectra of graphene. The data exhibits characteristic spectral features, reflecting specific properties of the unoccupied DOS of single-layer graphene. A comparison of spectra of single-layer, bilayer and FLG samples illustrates the rapid evolution of the electronic states from those of a truly two-dimensional systems, towards those of bulk graphite.

\section*{Acknowledgements}
We are grateful to Professor E. Cazzanelli for useful discussions. Part of this work was performed at the Swiss Light Source, Paul Scherrer Institut, Switzerland. This research project has been supported by the European Commission under the 6th Framework Programme through the Key Action: Strengthening the European Research Area, Research Infrastructures. Contract n¡: R113-CT-2004-506008. 

C.O.G., J.C.M., G. B., and A.Z. were supported by the Director, Office of Energy Research, Office of Basic Energy Sciences, Materials Sciences and Engineering Division, of the U.S. Department of Energy under Contract No. DE-AC02-05CH11231, via the sp2-bonded nanostructures program.

\newpage

\section*{Figure captions}
\textbf{Fig.1} C(K)-edge photoabsorption spectra of (from the bottom): graphene, bilayer graphene and FLG samples. Dashed lines show the C1s -$\pi*$ and C1s-$\sigma*$ transitions.

\textbf{Fig.2} C(K)-edge photoabsorption spectra of (from the bottom): graphene, bilayer graphene and FLG samples. Dashed lines show the C1s -$\pi*$ and C1s-$\sigma*$ transitions.

\textbf{Fig.3} Expanded C(K)-edge photoabsorption spectra of graphene and bilayer graphene extracted from Fig. 2. The inset shows corresponding DOS calculations, extracted from Ref. [9], in the energy region (0.0$\div$5.0) eV above $E_{F}$. Arrows show the double structure of the $\pi*$ resonance in graphene. The separation between the two features in the theoretical DOS of 1L
is about 0.8 eV, while the experimental separation is between 1.2 and
1.8 eV for our samples.

\textbf{Fig.4} C(K)-edge photoabsorption spectra of graphene taken from sample S2. In the inset, the Raman spectrum of graphene at 514 nm  showing the single component of the 2D peak.

\end{document}